\documentclass[aps,pra,twocolumn,groupedaddress,showpacs]{revtex4-1}
\usepackage{graphicx}
\usepackage{amsmath}
\usepackage{textcomp}
\usepackage{mathrsfs}
\usepackage{amsfonts}
\usepackage{mathtools}
\usepackage{tikz} 
\usetikzlibrary{positioning}
\usetikzlibrary{shapes,arrows,positioning,automata,backgrounds,calc,er,patterns}
\usepackage{tikz-feynman}
\tikzfeynmanset{compat=1.0.0}

\begin{document}

%new commands
\newcommand{\vett}[1]{\textbf{#1}}
\newcommand{\uvett}[1]{\hat{\textbf{#1}}}
\newcommand{\fieldE}[2]{E(\vett{#1},#2)}
\newcommand{\fieldA}[2]{A(\vett{#1},#2)}

\newcommand{\beq}{\begin{equation}}
\newcommand{\eeq}{\end{equation}}
\newcommand{\barr}{\begin{eqnarray}}
\newcommand{\earr}{\end{eqnarray}}
\newcommand{\bseq}{\begin{subequations}}
\newcommand{\eseq}{\end{subequations}}
\newcommand{\bal}{\begin{align}}
\newcommand{\eal}{\end{align}}
\newcommand{\ket}[1]{|#1\rangle}
\newcommand{\bra}[1]{\langle #1|}
\newcommand{\expectation}[3]{\langle #1|#2|#3 \rangle}
\newcommand{\braket}[2]{\langle #1|#2\rangle}

\title{Spectral Index of OAM-Carrying Ultrafast Localized Pulses}

\author{Souvik Agasti$^{1}$}
\author{Marco Ornigotti$^{1}$}
\email{marco.ornigotti@tuni.fi}
\affiliation{$^1$ Faculty of Engineering and Natural Sciences, Photonics Laboratory, Tampere University, 33720, Finland}

\begin{abstract}
We investigate the spectral degree of freedom of OAM-carrying localized waves, and its influence on their transverse intensity distribution. In particular, we focus our attention on two different families of spectra, namely exponentially decaying spectra, which are very tightly connected to fundamental X-waves, and X-waves with a Bessel spectrum. For each class we show how it is possible, by suitable manipulating their spectrum, to structure their transverse intensity distribution, thus creating a radial structure similar to that of Laguerre-Gaussian beams. To complete our analysis, we investigate the impact of the two spectral functions on linear and angular momentum of such localized waves.
\end{abstract}

\maketitle
\section{Introduction }

Since the invention of the laser in the 1960s by Maiman \citep{MAIMAN1960_laser}, optical beams have been the subject of extensive studies, mainly because of their versatility and potential for ground-breaking applications in the field of atomic and optical physics,such as material processing \citep{Material_processing}, imaging and spectroscopy \citep{spectroscopy_imaging, microscopy_imaging}, optical communications \citep{Fiber_Optic_Communication_agarwal, AVS_Quantum_Science} and fundamental research \citep{book_Ultrafast_Optics, book_Laser_Pulse_Gen_Appl}, to name a few. Recently, moreover, optical beams have been employed as an efficient mean to encode and transmit classical \citep{encode_info_OAM, marco_ultrashort_pulse} information, and as a viable solution to realise high-dimensional quantum information and communication protocols  \citep{quantum_qudit}, and error free quantum communication \citep{qudit_communication, qudit_entanglement_robert}. The key of success of optical beams in such context is mostly due to the possibility to structure, almost at will, their various degrees of freedom, a feature which allows great versatility. Achieving full control on degrees of freedom such as polarisation, orbital angular momentum (OAM), and spatial modes, understand their interplay, and exploit them for various applications is the vibrant subject of current research in the field \citep{Roadmap_structured_light}.

Recently, a lot of effort has been made to understand how structured light can be generalised to optical pulses as well. For example, Abourrady and co-workers have investigated the possibility of suitably structuring the dispersion relation of optical pulses, in order to suitably represent, and structure, monochromatic and polychromatic light fields \citep{space_optics, space_optics_nature}.
On a seemingly different ground, nondiffracting pulses, i.e., solutions of Maxwell's equations that are affected by neither dispersion nor diffraction during propagation, also offer many possibilities in terms of bringing structured light fields to the polychromatic domain \citep{book_Localized_Waves}.  Their most well-known representative is  the X-wave.  Firstly introduced in acoustics \citep{Nondiffracting_X_waves, Experimental_Nondiffracting_X_waves}, hey have, in fact, provided many interesting results in several different areas of physics, like nonlinear optics \citep{nonlinear_optics_INPROCEEDINGS}, condensed matter \citep{Conti_Ultracold_Gas}, quantum optics \citep{Quantum_X_Waves, Marco_Squeezing_oam, Marco_Quantum_X_waves}, integrated optics \citep{Nonlinear_Waveguide_X_Wave, X_Waves_Photonic_Lattices} and optical communications \citep{X_wave_communications}. Recently, moreover, X-waves carrying OAM have been studied theoretically, and an interesting interplay between their temporal dynamics and OAM characteristic has been reported \citep{marco_read1, marco_ultrashort_pulse, marco_polarized_pulse}.

Due to their intrinsic resilience to external perturbations, nondiffracting pulses, and X-waves, in particular, represent a very interesting platform for free-space classical and quantum communication. However, X-waves are typically only characterised by a single index, i.e., their OAM, and therefore do not possess a characteristic radial structure, such as Laguerre-Gaussian beams, for example, that would give them the possibility to be used for high-dimensional quantum information and communication protocols. Since X-waves are essentially polychromatic superpositions of Bessel beams, one possible way to overcome this problem would be to directly generalise the results concerning Bessel beams with two indices \citep{marco_Bessel_monochroma} to the domain of optical pulses. In doing so, however, one would have to deal with superposition of X-waves, rather than single X-waves, as Bessel beams with two indices exist only as superposition of Bessel beams.

Alternatively, one could investigate the possibility to exploit the peculiar structure of nondiffracting beams, i.e., their spatio-temporal correlation, to induce a radial structure into them, by simply shaping their frequency spectrum. This is the approach we will follow in this work. In particular,  we aim at introducing  a new degree of freedom, which we call \emph{spectral index}, which indices, as we will show,  a radial structure into X-waves similar of that of Laguerre-Gaussian beams.
 Here, we focus our attention on two different classes of X-waves, namely fundamental X-waves, possessing exponentially decaying spectra, and X-waves with Bessel spectrum, and investigate how the spectral properties of each class shapes the transverse structure of the pulse. While the former class is very well-known and largely studied, the latter has not been investigated yet, and we introduce it for the first time. To make our analysis complete, we also investigate how their vector properties, namely linear and angular momentum, are affected by such spectral distributions.

This work is organised as follows: in Sect. II we briefly recall the definition of X-waves, both in their scalar and vector form, and we give the general expressions of their electric and magnetic fields. Section III is then dedicated to fundamental X-waves. There, we show how changing their spectral order results in the appearance of a transverse spatial structure similar to that of Laguerre-Gaussian beams. We then repeat the same kind of analysis in Sect. IV for X-waves with Bessel spectrum. A brief discussion on the physical meaning of the spectral index for both families of localized waves is given in Sect. V. Section VI is then dedicated to the analysis of their linear and angular momentum characteristics, within the assumption of paraxial propagation. Finally, conclusions are drawn in Sect. VII.

\section{From Nondiffracting Beams to Nondiffracting Pulses}
We start our analysis by considering  a scalar, monochromatic solution of the Helmholtz equation
\begin{equation} \label{Helmholtz Equation k space}
(\nabla^2 +k^2) \psi(\textbf{r},k)=0,
\end{equation}
where $k = 2\pi/\lambda$ is the wave vector in vacuum. In a cylindrical reference frame $\{R,\theta, z\}$, the solution can be given in term of Bessel beams
\begin{equation}\label{bessel fn}
\psi(\textbf{r},k)=J_m(k R\sin \vartheta_0)e^{im\theta} e^{ikz\cos\vartheta_0},
\end{equation}
where  $R = \sqrt{x^2 + y^2}$, $\theta = \arctan(y/x)$, $J_m(x)$ is the Bessel function of the first kind,  $m$ is the OAM index, and $\vartheta_0$ is the Bessel cone angle which is the beam’s characteristic parameter \citep{Diffraction-Free_Beams}. Notice, that $\psi(\vett{r},k)$ is propagation invariant, as the $z-$dependence is contained only in the plane-wave term $\exp{(ikz\cos\vartheta_0)}$, and thus the intensity distribution of a Bessel beam does not change during propagation, i.e., $\partial_z|\psi(\vett{r},k)|^2=0$. With this solution at hand, we can construct an exact solution of the wave equation
\begin{equation}\label{Helmholtz Equation t space}
\left(\nabla^2 -\frac{1}{c^2}\partial_t^2\right) \phi(\textbf{r},t)=0,
\end{equation}
as a polychromatic superposition of monochromatic solution in the following way
\begin{equation}\label{scalar field}
\phi(\textbf{r},t)=  \int \mathrm{d}k\, g(k)\,e^{-ickt}\psi(\textbf{r},k),
\end{equation}
where $g(k)$ is an arbitrary spectral function. If $\phi(\vett{r},t)$ is constructed from Bessel beams, then the $z-$ and $t-$ dependence of each spectral component will be of the kind $\exp{[i k(z\cos\vartheta_0-c t)]}$, which is the distinctive characteristic of localized waves \citep{book_Localized_Waves}. {For nondiffracting pulses, in fact, regardless of the explicit form of $g(k)$, it can be demonstrated, that they are propagation invariant with respect to both propagation direction and time, or, alternatively, with respect to the co-moving propagation direction $\zeta=z\cos\vartheta_0-ct$, i.e., $\partial_{\zeta}|\phi(\vett{r},t)|^2=0$.

This scalar solution can be used to generate exact vectorial solutions of Maxwell's equation, using the method of Hertz's potentials \citep{Electromagnetic_Theory_book}.  In fact, if we define the Hertz potential
\beq\label{hertz potential}
\boldsymbol\Pi(\vett{r},t) = \int\,dk\,g(k)\,\psi(\vett{r},k)\,e^{-ickt}\,\uvett{u},
\eeq
the (TM) vector electric and magnetic fields are defined as
\bseq \label{TE vector eq}
\begin{align}
\mathbf{E}(\textbf{r},t) &= \nabla\times \nabla \times \mathbf{\Pi}(\textbf{r},t), \\
\mathbf{B}(\textbf{r},t) &= \frac{1}{c^2}\,\frac{\partial}{\partial t} [ \nabla \times \mathbf{\Pi}(\textbf{r},t) ].
\end{align} 
\eseq
In the case of TE fields, the roles of the electric and magnetic fields in the above equation are exchanged (together with an extra minus sign between the TM magnetic field and the TE electric field).

Due to the linearity of Eq. \eqref{hertz potential}, we can first calculate the vector electric and magnetic fields of the monochromatic components of the nondiffracting pulse, and then integrate over $k$. In doing this, then, we first define the monochromatic Hertz potential
\begin{equation}\label{hertz potential monochromatic}
\mathbf{P}(\textbf{r},t,k) = \psi(\vett{r},k)e^{-i c k t}\,\uvett{u},
\end{equation}
then define the monochromatic vector electric ($\mathcal{E}(\vett{r},t;k)$) and magnetic ($\mathcal{B}(\vett{r},t;k)$) in a similar manner to Eqs. \eqref{TE vector eq}, and, lastly, we can write the vector electric and magnetic fields of the nondiffracting pulse as a function of their monochromatic counterparts as follows:
\bseq\label{fieldMon}
\begin{align}
\mathbf{E}(\textbf{r},t) &= \int \mathrm{d}k  \, g(k) \, \mathcal{E}(\textbf{r},t,k),\\
\mathbf{B}(\textbf{r},t) &= \int \mathrm{d}k  \, g(k) \, \mathcal{B}(\textbf{r},t,k).
\end{align}
\eseq
The explicit expressions of $\mathcal{E}(\vett{r},t;k)$ and $\mathcal{B}(\vett{r},t;k)$ for Bessel beams in given in Appendix A. This way of calculating the electric and magnetic fields of nondiffracting pulses is quite useful when dealing with vector quantities, such as the linear and angular momentum, as the well-known results for monochormatic beams need simply to be integrated, with the corresponding spectral function characterising the different nondiffracting pulses.

Notice, moreover, that vectorialising a scalar solution of the Helmholtz equation with the method of Hertz potentials also implicitly requires a suitable gauge transformation} \citep{bessel_quanta} to guarantee the transversality of the vector fields.

In the remaining of this manuscript, we will assume $\uvett{u}=\uvett{z}$ for simpicity, and we will concentrate our attention to two spectral functions in particular, namely the generalised exponentially decaying spectrum 
\beq\label{expDecaying}
g_{exp}(k)=\Theta(k)\,k^n\,e^{-\alpha k},
\eeq
(where $n\in\mathbb{N}$ defines the order of the spectrum,  $\alpha>0$ accounts for the width of the spectrum, and $\Theta(k)$ is the Heaviside step function \citep{book_NIST}) and the decaying Bessel spectrum 
\beq\label{besselDecaying}
g_{bessel}(k)=\Theta(k)\,\text{J}_{2m}(2a\sqrt{k})e^{-\alpha k},
\eeq 
where $m$ is the OAM index of the momochromatic Bessel beam associated with this spectrum, and  $a\in\mathbb{R}$ (which has the dimensions of  $\sqrt{\text{length}}$) is a free parameter determining the width of the spectral function $g_{bessel}(k)$.

The spectrum in Eq. \eqref{besselDecaying} is closely related to the so-called spinor wave focus modes \cite{hillion1}, a variant of which has been proposed to be sustainable in optical fibers \cite{hillion2}. Understanding the properties of X-waves with Bessel spectrum in free-space would therefore be important to identify the key ingredients to structure light pulses in such a way, that they could efficiently be coupled with optical fibers and, more in general, integrated systems.

It is worth noticing, that although at a first glance it might seem that X-waves with Bessel spectrum only look like a theoretical curiosity, with almost no possibility to be realised experimentally, the recent progress in manipulating the spectrum of a laser pulse (see, for example, Ref. \cite{airyPulse} where the spectrum of a pulse is shaped in the form of an Airy function) suggests that the experimental realisation of X-waves with Bessel spectrum is not so impossible as one might think.

Section III will be dedicated to investigate the properties, both in the scalar and vector case, of $g_{exp}(k)$, its connection with fundamental X-waves, and how it is possible, by suitably choosing the value of $n$, to induce a radial structure, similar to that of Laguerre-Gaussian beams \citep{book_Principles_lasers}. The same kind of analysis will be conducted in Sect. IV for the case of $g_{bessel}(k)$, where a connection between the parameter $a$ and the radial structure of the pulse will be established.

\begin{figure}[t!]
\centering
\includegraphics[width=1\linewidth]{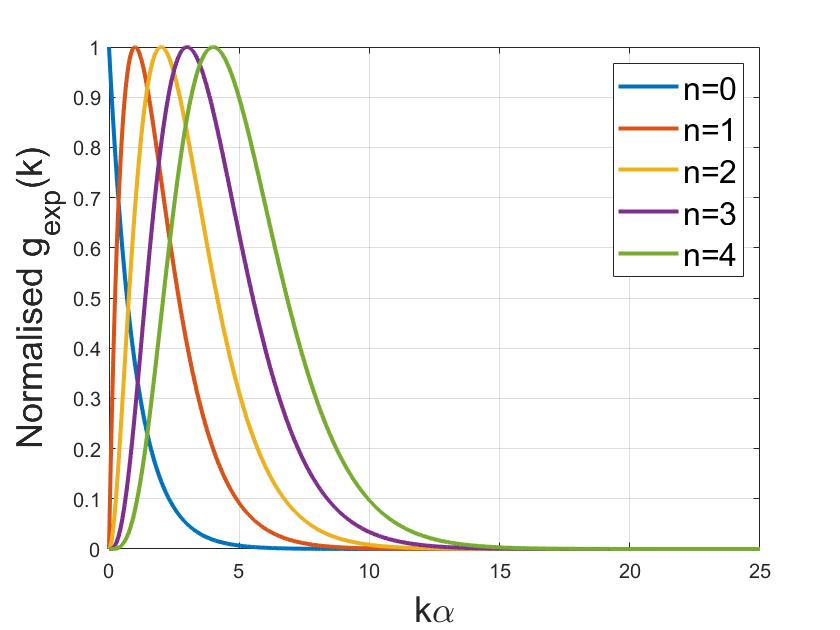}
\caption{Plot of the exponentially decaying spectral function $g_{exp}(k)$, as given by Eq. \eqref{expDecaying}, for different values of the spectral index $n$. As it can be seen, as $n$ grows, the spectrum shifts towards higher frequencies, while its FWHM grows according to Eq. \eqref{spectralFWHM}.} \label{k_vs_gk_expo}
\end{figure}

\section{Exponentially Decaying Spectrum}
\begin{figure*}[t!]
\centering
\includegraphics[width=1\linewidth]{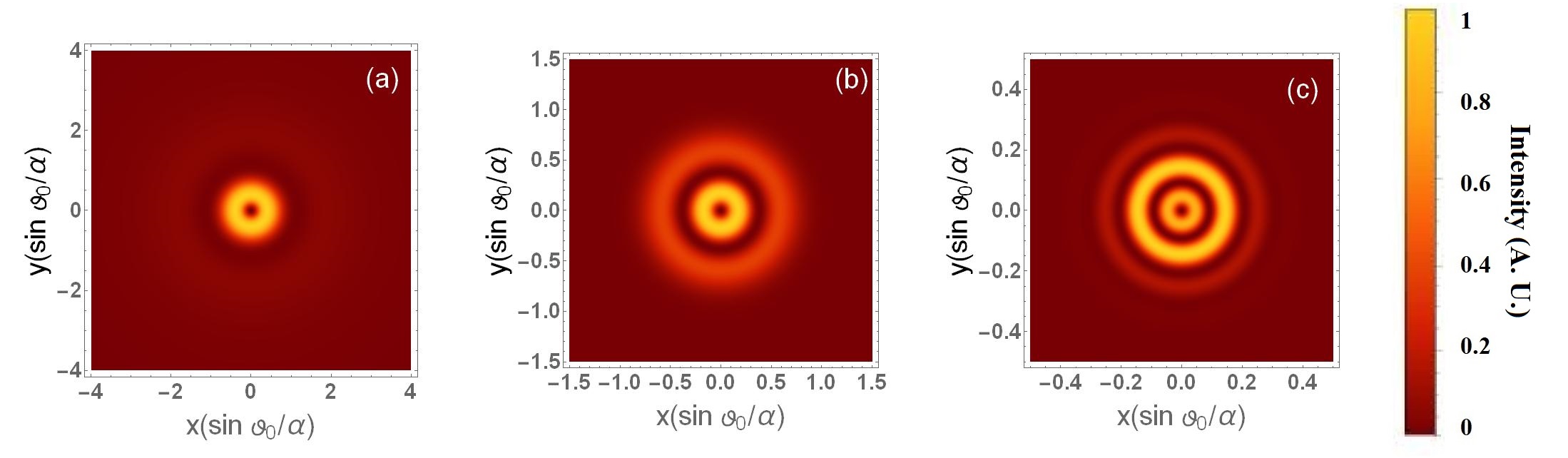}
\caption{Plot of intensity distribution of the radial component $E_r$ of the electric field of exponentially decaying nondiffracting pulses in the plane $\zeta = 0$ for different values of the spectral order parameter $n$:  (a) $n=0$, corresponding to fundamental X-wave carrying OAM. (b) $n=5$,  and (c) $n=30$. Notice, that as $n$ grows, the radial structure of $E_r$ changes, and new rings appear around the starting OAM ring, thus building a structure similar to Laguerre-Gaussian modes. For all the plots above, $m=2$ has been used as a reference OAM value. Moreover, $\vartheta_0 = 0.01$, and $\alpha= 1$ have been assumed.} \label{EB_field_expo(m=2,n=0,5,30)}
\end{figure*}
Let us consider the exponentially decaying spectrum $g_{exp}(k)$ given by Eq. \eqref{expDecaying}, and remember, that $n\in\mathbb{N}$ is a constant representing the spectral order. As it can be seen from Fig. \ref{k_vs_gk_expo}, as $n$ increases, the spectrum acquires a bell-like shape, which shifts to higher frequencies with increasing $n$. In particular, we observe, that as $n$ grows, the maximum shifts linearly with $n$ towards higher frequencies, and the full width half maximum (FWHM) of the spectrum changes according to 
\beq\label{spectralFWHM}
\Delta k_{FWHM} = -\frac{2n}{\alpha} \text{W}_0\left(-\frac{\alpha}{n\,2^{1/n}}\right),
\eeq
where $W_0(x)$ is the Lambert W-function \citep{book_NIST}. Notice, that $\Delta k_{FWHM}$ grows quite slowly with $n$, and, to first approximation, it remains roughly constant. Moreover, it asymptotically tends to a constant value, for $n\rightarrow\infty$. If we substitute Eq. \eqref{expDecaying}  into Eq. \eqref{scalar field} and use the relation 6.621.1 of Ref. \citep{table_of_integrals_book}, namely
\barr\label{integral_expdcy}
\int_0^\infty &\mathrm{d}x& \,e^{-\gamma x}\, J_\nu (\beta x)\, x^{\mu-1} = \frac{(\beta/2\gamma)^\nu \Gamma(\nu+\mu)}{\gamma^\mu \Gamma(\nu+1)} \nonumber\\
 &&F\left(\mu,\nu; -\beta^2/\gamma^2  \right),
\earr
where $F(\mu,\nu;x)=\,_2F_1((\mu+\nu)/2,(\mu+\nu+1)/2;\nu+1;x)$ is the Gauss hypergeometric function. With this relation, the scalar pulse $\phi(\vett{r},t)$, known in literature as fundamental X-wave, has the following form
\begin{equation}\label{fundamentalX}
\phi_m^{(n)} (\textbf{r},t) = C_{m,n}\,e^{im\theta}\,\rho^m\,F(m,n,-\rho^2/\xi^2),
\end{equation}
where $\rho=R\sin\vartheta_0$,  $ \xi = (\alpha-i\zeta)$ and $C_{m,n}=(m+n)!/(2^m \xi^{m+n+1}m!)$. Using the results from Appendix A for the vector fields, together with Eq. \eqref{integral_expdcy} above, we can write the expression for the vector (TM) electric field $\vett{E}_{TM}$ in cylindrical coordinates $\{R,\theta,\zeta\}$ and cylindrical reference frame $\{\uvett{R},\hat{\boldsymbol\theta},\uvett{z}\}$ of exponentially decaying localised waves as follows:
\bseq
\begin{align}
E_R(\vett{r},t) &= e^{im\theta}\Bigg[C_{m-1,n}\,\rho^{m-1}F\left(m-1,n;-\frac{\rho^2}{\xi^2}\right)\nonumber\\
&-C_{m+1,n}\,\rho^{m+1}\,F\left(m+1,n;-\frac{\rho^2}{\xi^2}\right)\Bigg],\\
E_{\theta}(\vett{r},t) &=\left(\frac{m\,C_{m,n-1}}{\sin\vartheta_0}\right)e^{im\theta}\,\rho^m\,F\left(m,n-1;-\frac{\rho^2}{\xi^2}\right),\\
E_z(\vett{r},t) &= \tan\vartheta_0\,\phi_m^{(n)}(\vett{r},t).
\end{align}
\eseq
Similar expressions can be derived for the (TM) magnetic vector field and vector potential as well, using the results given in Appendix A.

\subsection{Connection to Generalised X-Waves}
The fundamental X-wave $\phi_m^{(n)}(\vett{r},t)$ is very closely related to the so-called generalised X-waves \citep{book_Localized_Waves, marco_ultrashort_pulse}. While  fundamental X-waves  are generated by a purely exponentially decaying spectrum, like the one given by Eq. \eqref{expDecaying}, generalised X-waves are instead constructed starting from the following spectrum \citep{Orthogonal_X_wave} 
\beq\label{fundamental}
g_{gen}(k)=\Theta(k)\,C_n(k)\,\text{L}_n^{(1)}(\alpha k)e^{-\alpha k},
\eeq
where $\text{L}_n^{(1)}(x)$ are associated Laguerre polynomials of the first kind \citep{book_NIST} and $C_n(k)$ is a suitable normalisation constant. It is not difficult to see, using the definition of associated Laguerre polynomials, that the two spectra are connected by the following relation
\beq\label{genSpectrum}
g_{gen}(k)=\frac{\tilde{C}_n(k)}{k}\frac{d^n}{dk^n}\left(k\,g_{exp}(k)\right),
\eeq
i.e., the spectrum of generalised X-waves is obtained from fundamental ones by successive derivation. 

It is worth noticing, that the equation above implicitly constraints the the spectral index $n$ of fundamental X-waves to be an integer. To understand this, let us notice, that the expression above is tightly connected to the generating function of generalised Laguerre polynomials \cite{book_NIST}, i.e.

\beq\label{generatingFunction}
\text{L}_n^{(\alpha)}=\frac{x^{-\alpha}e^x}{n!}\frac{d^n}{dx^n}\left(e^{-x}x^{n+\alpha}\right).
\eeq
A direct comparisong between the above equarion and Eq. \eqref{genSpectrum} allows one to conclude that the spectral index $n$ of the spectrum $g_{exp}(k)$ appears as well as order of the derivative, and it is therefore constrained to be an integer. For this reason, rather than investigating the general case $n\in\mathbb{R}$, we limit ourselves to the particular case of the spectral index $n$ being an integer, so that the results of our analysis can be easily extended to generalised X-waves as well.

 Notice, moreover, that for $n=0$, both cases reproduce the standard result of X-waves \citep{book_Localized_Waves}. For the case $n\neq 0$, on the other hand, the two solutions have sensibly different spectra. In their classical context, however, X-waves are frequently only studied in the limit $n=0$, as most of their properties are independent on the shape of the spectrum, provided that the spectrum has an exponentially decaying component.

Contrary to the case of Laguerre-Gaussian (LG) and Hermite-Gaussian (HG) beams, which are characterised by two spatial indices \citep{angular_momentum_of_light}, which allow a borader control of their spatial structure, and can also be used to encode a larger amount of information in them \citep{robert_HG_LG_beam, quantum_qudit, qudit_entanglement_robert}, X-waves inherently possess only one spatial index, namely the OAM-index, which is inherited by their monochromatic Bessel components. However, X-waves possess a natural correlation between spatial and temporal degrees of freedom (i.e., the fact, that they are propagation invariant in the co-moving frame $\zeta=z\cos\vartheta_0-c t$), which can be exploited to induce a more richer spatial structure, by controlling their spectrum. This, as we will show in the remaining of this manuscript, can lead to the emergence of an ``effective" radial index, which can be used to further engineer the spatial profile of X-waves.
\subsection{Effective Radial Index of Fundamental X-Waves}
\begin{figure}[t!]
\centering
\includegraphics[width=1\linewidth]{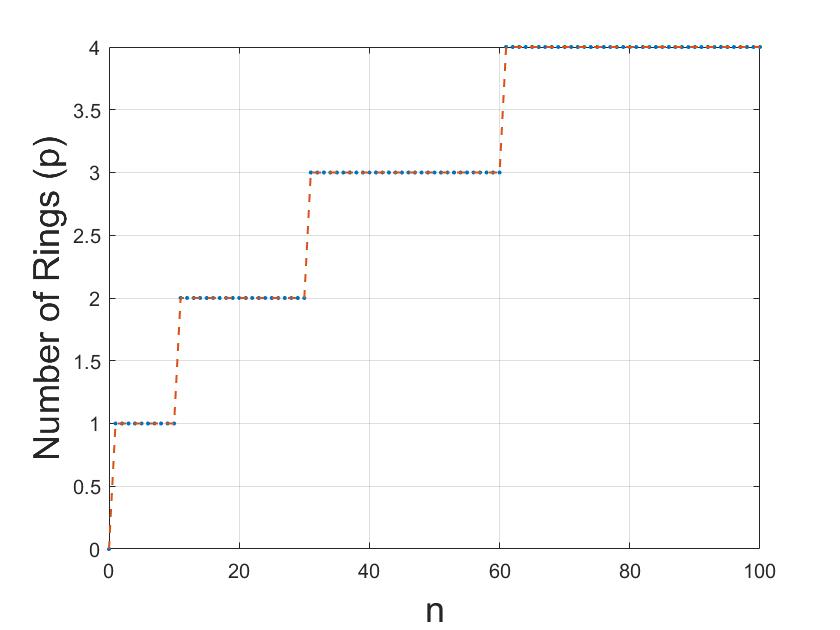}
\caption{ Plot of the number of visible rings (p) for different values of the spectral index $n$ (blue dots). To calculate the number of rings, the number of maxima appearing in the radial structure of one of the components of the electric field (in this case, the radial component $E_r$ was chosen, but the same analysis is valid for the other ones), as $n$ is varied from 0 to 100. Notice, that since we consider only paraxial pulses, we have limited our analysis to only the paraxial regione of the pulse, i.e., the region close to its center. The parameters of the electric field used to calculate the number of rings (p) are the same as Fig. \ref{EB_field_expo(m=2,n=0,5,30)}. A visual help (orange line) has been added, in order to facilitate the visualisation of the incremental step-changes in $p$. } \label{shift_exp}
\end{figure}
In this section we then consider the effect of the spectral index $n$ on the spatial structure of vector X-waves. As it appears clear from Fig. \ref{EB_field_expo(m=2,n=0,5,30)}, as $n$ increases, a richer spatial structure in the intensity distribution of their vector electric field (here represented by its radial component, but the same can be said for the other components too). In particular, from Fig. \ref{EB_field_expo(m=2,n=0,5,30)} it seems that the pulses acquire a radial structure very similar to that of LG beams. This result can be directly compared to previous ones concerning assigning a second index to monochromatic Bessel beams \citep{marco_Bessel_monochroma}. However, while in the case of Bessel beams with two indices, the second index, and therefore a LG-like radial structure, was realized by constructing suitable superpositions of Bessel beams with different transverse wave vectors \citep{marco_Bessel_monochroma} in this case, the additional radial structure is due to the form of the spectrum. In a sense, one could say, that by suitably superimposing Bessel beams of different frequencies, it is possible to induce a radial structure to them. To some extent, this is another manifestation of the spatio-temporal correlation typical of localized waves. \citep{book_Localized_Waves}.

To better understand the connection between spectral order and radial structure of fundamental X-waves, in Fig. \ref{shift_exp} we plot the number of rings appearing in the radial component of the vector electric field, as a function of the discrete spectral index $n$. There,  we observe that it is possible to establish a connection between the range of values of $n$ that allow a certain number or rings $p$ by means of the following recursive relation:
\beq\label{pIndex}
\mathcal{I}_n(p+1) \equiv \{ n\in\mathbb{N} /\, n_1 \leq n \leq n_2\}
\eeq
where $n_1=1+10(p-1)+\text{max}\{\mathcal{I}_n(p-2)\}$, $n_2=10p+\text{max}\{\mathcal{I}_n(p-1)\}$, and the following initial conditions are implicitly understood $\text{max}\{\mathcal{I}_n(-2)\}=\text{max}\{\mathcal{I}_n(-1)\}=\text{max}\{\mathcal{I}_n(0)\}=0$, and  $\mathcal{I}_n(p)$ stands for the interval of allowed values of $n$ for a given number of rings $p$. Notice, moreover, that within the same interval $\mathcal{I}_n(p)$ different values of $n$ correspond to the same number of rings, but a different ratio of intensity between them. In this way, therefore, not only it is possible to induce a richer spatial structure to the X-wave, but one can also control the relative weight between the various rings.

In analogy with LG-beams, where the $p$ index regulates the radial structure of the field mode, we then introduce a \emph{spectral} $p$-index, which introduces a richer radial structure in the intensity distribution of localized waves. Unlike LG-beams, however, the origin of this radial structure is not to be sought in a coupling of spatial dimensions due to the choice of coordinates, but rather to the spatio-temporal correlation typical of localized waves.

Notice, moreover, that as new rings appear in the transverse structure of the X-wave, a global shrinking of its intensity distribution is observed. This effect is another manifestation of the innate spatio-temporal correlation typical of localized waves, and it is mainly due to the fact, that the polychromatic superposition of Bessel beams given by Eq. \eqref{scalar field} can be interpreted as being the Hankel transform of the spectral distribution $g_{exp}(k)$. In this case, as $n$ grows, and the bell-shaped curve in Fig.  \ref{k_vs_gk_expo} shifts towards higher frequencies, the spectral FWHM also grows according to Eq. \eqref{spectralFWHM}, and therefore, the spatial size of the pulse shrinks, since it is directly connected, through Eq. \eqref{scalar field}, to its frequency spectrum.

This is the first result of our work. By suitably changing the form of the spectrum of fundamental X-waves, i.e., by varying the spectral index $n$, it is possible to induce a modification in the spatial intensity distribution of the X-waves itself, i.e., by introducing a radial structure similar to that of LG-beams. Given a single value of the spectral index $p$, which regulates, in analogy with LG-beams, the number of rings possessed by the X-waves, there exists an interval of allowed values of the index $n$ of the X-wave, that possess that number of rings. Within the same interval, different values of $n$ tune the ratio in intensity between the various rings.

\section{BESSEL SPECTRUM}
\begin{figure}[t!]
\centering
\includegraphics[width=1\linewidth]{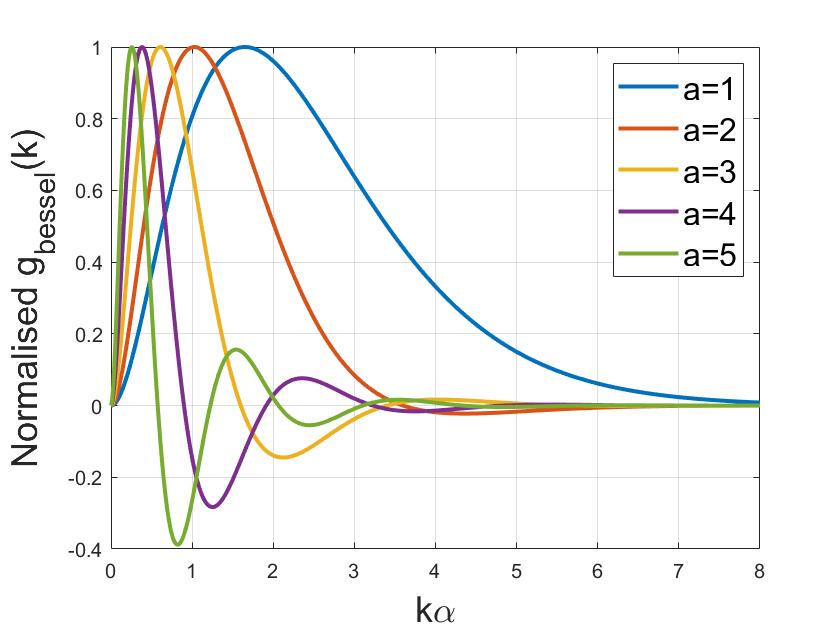}
\caption{Plot of the exponentially decaying Bessel spectrum $g_{bessel}(k)$, as given by Eq. \eqref{besselDecaying}, for different values of the continuous spectral index $a$. As it can be seen, as $a$ grwos, the modulation induced by the Bessel function $\text{J}_{2m}(a\sqrt{k})$ introduces spectral regions of negative weight (i.e., a phase shift of $\pi$ with respect to the other regions of the spectrum), which, in turn, correspond to a richer structure in the transverse intensity distribution. The appearance of oscillations in the spectrum, in fact, is directly connected to the appearance of rings in the transverse structure of the pulse. Here, $m=2$ has been assumed, to match the aprameters used in Fig. \ref{ER_field_bessel(m=2,a=1,5,10)}.} \label{k_vs_gk_bessel}
\end{figure}
\begin{figure*}[t!]
\centering
\includegraphics[width=1\linewidth]{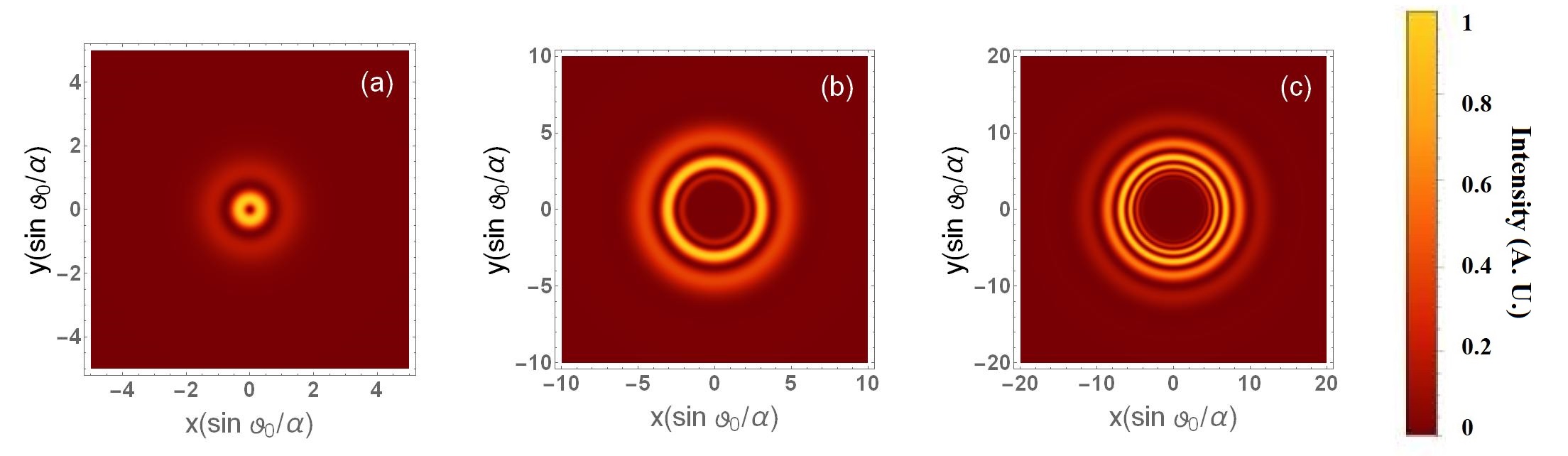}
\caption{ Plot of intensity distribution of the radial component $E_r$ of the electric field of X-waves with Bessel spectrum in the plane $\zeta = 0$ for different values of the spectral order parameter $a$:  (a) $a=1$, which has a very strong resemblance with a fundamental X-wave carrying OAM, depicted in Fig. \ref{EB_field_expo(m=2,n=0,5,30)}(a) . (b) $a=5$,  and (c) $a=10$. Notice, that as $a$ grows, the radial structure of $E_r$ changes, and new rings appear around the starting OAM ring, thus building a structure similar to Laguerre-Gaussian modes. For all the plots above, $m=2$ has been used as a reference OAM value. Moreover, $\vartheta_0 = 0.01$, and $\alpha= 1$ have been assumed.} \label{ER_field_bessel(m=2,a=1,5,10)}
\end{figure*}
Fundamental X-waves are not the only localized wave to possess this feature. Similar results as those obtained above can also be seen in a different class of X-waves, possessing a Bessel spectrum, as the one give in Eq. \eqref{besselDecaying}. Contrary to fundamental X-waves, the spectrum of X-waves with Bessel spectrum is regulated by a \emph{continuous}, rather than a discrete, parameter. The form of this spectrum for several values of the free parameter $a\in\mathbb{R}$ is given in Fig. \ref{k_vs_gk_bessel}. As it can be seen, as $a$ grows bigger, while the exponential tail suppresses the oscillations of the Bessel function for large $k$, for small $k$ new oscillations appear, because of the decreasing size of the effective width of the spectrum, proportional to $1/a$. 

Substituting this spectral function into Eq. \eqref{scalar field}, and using Eq. 6.644 of Ref. \citep{table_of_integrals_book}, namely
\beq\label{integral_besseldcy}
\int_0^\infty \mathrm{d}x e^{-\gamma x} J_{2\nu} (2a \sqrt{x}) J_\nu (\beta x)  = \frac{e^{-\Xi\alpha}\text{J}_m\left(\Xi\beta\right)}{\sqrt{\gamma^2+\beta^2}},
\eeq
where $\Xi=a^2/(\gamma^2+\beta^2)$, we obtain the scalar field for localized waves with Bessel spectrum as follows
\beq\label{besselX}
\phi_m(\textbf{r},t;a) =\frac{e^{im\theta}}{\xi\,\sqrt{R(\rho)}}e^{-\frac{a^2}{\xi\,R(\rho)}}\,\text{J}_m\left(\frac{a^2\rho}{\xi^2\,R(\rho)}\right),
\eeq
where $R(\rho)=1+\rho^2/\xi^2$.
Contrary to the previous case, however, for  X-waves with Bessel spectrum we cannot use Eqs. \eqref{fieldMon} to calculate the electric and magnetic vector fields, since the spectrum $g_{bessel}(k)$ requires the index of the spatial and spectral Bessel functions to be linked with each other. By first vectorialising the monochomatic solution and then integrating with respect to $k$, in fact, this constrain will not be satisfied anymore, as the derivatives with respect to $R$ contained in the curl operator will change the index of the spatial Bessel function, thus rendering Eq. \eqref{integral_besseldcy} not applicable anymore. In this case, therefore, we can only apply directly Eqs. \eqref{TE vector eq}, which result in the following electric field components:
\bseq
\begin{align}
E_R(\vett{r},t;a) &= -\frac{i}{2}\sin2\vartheta_0\,\frac{\partial^2\phi(\vett{r},t;a)}{\partial\xi\partial\rho},\\
E_{\theta}(\vett{r},t;a) &= \frac{m\sin2\vartheta_0}{2\rho}\,\frac{\partial\phi(\vett{r},t;a)}{\partial\xi},\\
E_z(\vett{r},t;a) &=-\sin^2\vartheta_0\nabla^2_{\rho}\phi(\vett{r},t;a),
\end{align}
\eseq
where $\nabla^2_{\rho}=\left[\frac{1}{\rho}\frac{\partial}{\partial\rho}\left(\rho\frac{\partial}{\partial\rho}\right)+\frac{m^2}{\rho}\right]$. The magnetic field can be analogously calculated.

\begin{figure}[t!]
\centering
\includegraphics[width=1\linewidth]{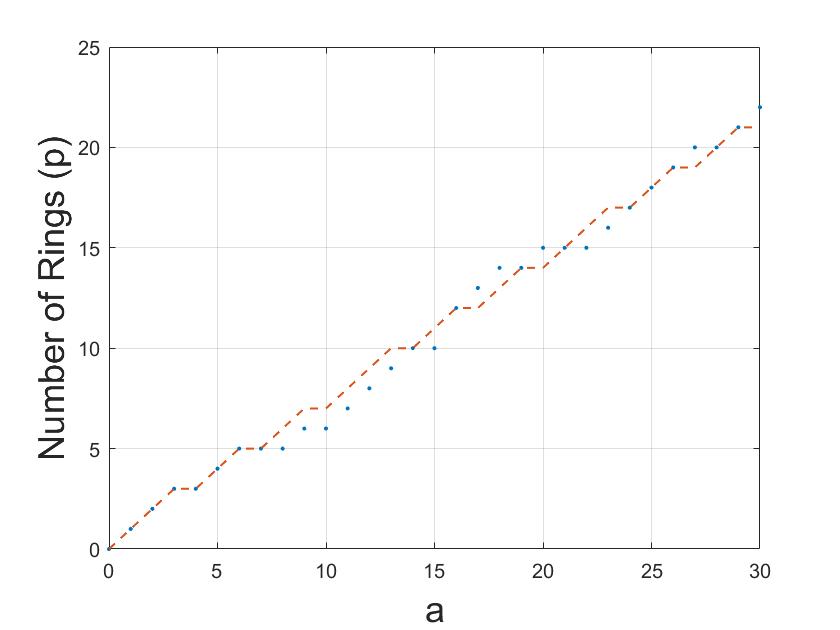}
\caption{ Plot of the number of visible rings for different (discrete) values of the spectral index $a$ (blue dots). To calculate the number of rings, a similar algorithm to the one used for fundamental X-waves has been employed. The parameters of the electric field used to calculate the number of rings (p) are tha same as Fig. \ref{ER_field_bessel(m=2,a=1,5,10)}. A visual help (orange line) has been added, in order to facilitate the visualisation of the incremental step-changes in $p$.} \label{exp_bess}
\end{figure}
The radial component of the electric field defined above is plotted in Fig. \ref{ER_field_bessel(m=2,a=1,5,10)} as a function of the free parameter $a$, which, in this case, plays the role of spectral order parameter (that was $n$, for fundamental X-waves).
\begin{figure*}[t!]
\centering
\includegraphics[width=1\linewidth]{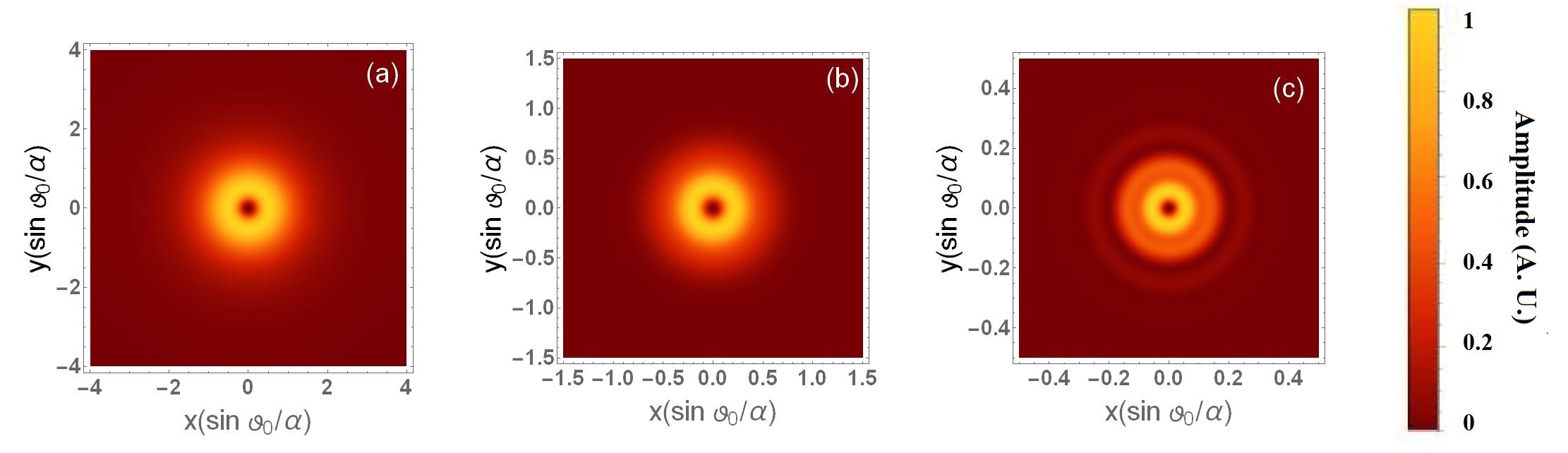}
\caption{Plot of intensity distribution of the longitudinal component $\mathcal{P}_z(\vett{r})$ of the linear momentum (Poynting vector) of fundamental X-waves in the plane $\zeta = 0$ for (a) $n=0$, (b) $n=5$, and (c) $n=30$. All other parameters remain same as Fig. \ref{EB_field_expo(m=2,n=0,5,30)}. } \label{linear_momenta_expo(m=2,n=0,5,30)}
\end{figure*}

As it can be seen by comparing Fig. \ref{EB_field_expo(m=2,n=0,5,30)} and \ref{ER_field_bessel(m=2,a=1,5,10)}, even in this case, as the spectral order parameter $a$ varies, a richer spatial structure is induced in the electric field. Because of the different generating spectrum, however, the radial structure of X-waves with Bessel spectrum is fundamentally different from the one of fundamental X-waves. In fact, while in the latter case the extra rings, as $n$ grows, appear outside the central structure of the pulse (loosely speaking, the are dragged in from the infinity), in the former case, instead, the rings are created within the waist of the pulse (i.e., pushed out from the center of the beam), thus creating a more dense pattern. This, essentially, is due to the different spatial form of the two pulses. For fundamental X-waves, the spatial behaviour is governed by the Gauss' hypergeometric function, while for X-waves with Bessel spectrum it is, essentially, due to a Bessel function [see Eqs. \eqref{fundamentalX} and \eqref{besselX}].

Notice, however, that, unlike the case of exponentially decaying spectrum [Fig. \ref{shift_exp}], in the case of a Bessel spectrum the number of rings $(p)$ changes linearly while changing spectral index,  $a$, as it is depicted in Fig. \ref{exp_bess}. A significant difference with the case discussed in the previous section is, that in this case the spectral order parameter $a$ is a continuous, rather than discrete. However, given the piecewise linear relation between $a$ and the number of rings appearing in the transverse structure of the field, it is not difficult to define a discrete index starting from $a$, which can be understood as a radial index for the spatial profile of localized waves, namely
\beq\label{pIndexBessel}
p=\left\lceil c_1\,a\right\rceil,
\eeq 
where $\lceil\,x\,\rceil$ is the ceiling function, which rounds its argument to the next integer value, and $c_1= 0.7$, as can be extrapolated from Fig. \ref{exp_bess} by linear fitting.
\begin{figure*}[t!]
\centering
\includegraphics[width=1\linewidth]{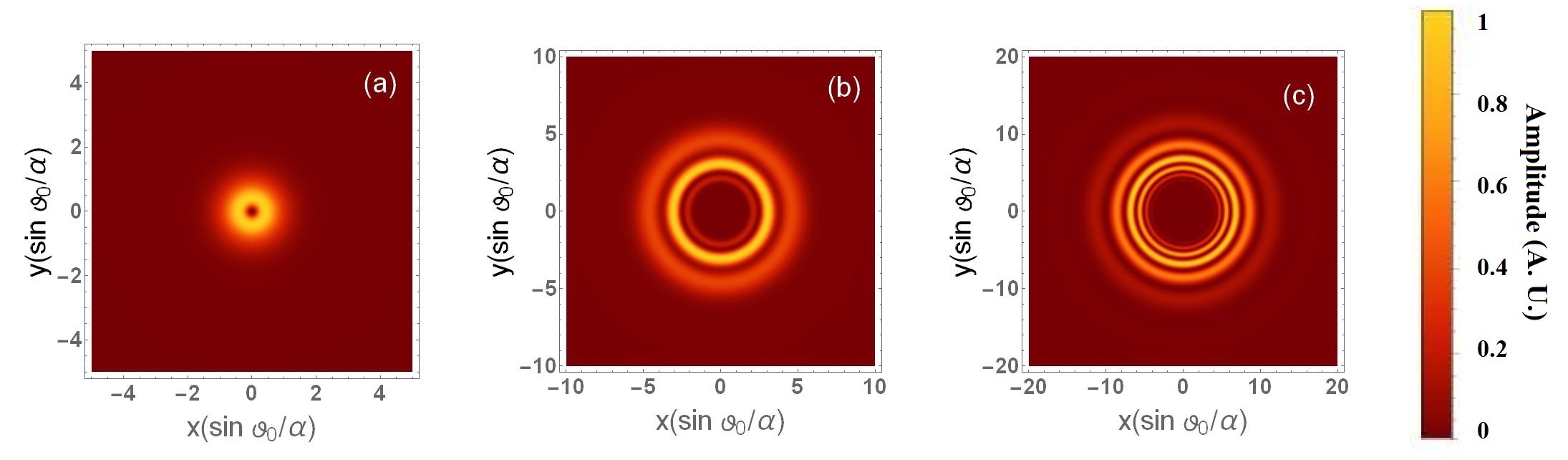}
\caption{Plot of intensity distribution of the longitudinal component $\mathcal{P}_z(\vett{r})$ of the linear momentum (Poynting vector) of X-waves with Bessel spectrum in the plane $ \zeta=0.$ for  (a) $a=1$, (b) $a=5$, and (c) $a=10$. All other parameters remain same as Fig. \ref{ER_field_bessel(m=2,a=1,5,10)}. }\label{bessel_fn_P(m=2,a=1,5,10)}
\end{figure*}
\section{On the Nature of These New Radial Indices}
It is important to discuss the nature of the spectral indices introduced for both families of X-waves above. To do this, let us first recall, that for the case of LG-beams $LG_p^m(\vett{r})$, for example, both radial ($p$) and azimuthal ($m$) index are \emph{proper} indices, in the sense, that it is possible to associate an orthogonality relation to each of those indices, i.e., that (abusing the quantum formalism) $\langle LG_{q,n}|LG_{p,m}\rangle=\delta_{m,n}\delta_{p,q}$.

This, however, is not true for the two cases presented above. For fundamental X-waves, in fact, the scalar product of two pulses with different spectral index gives
\barr
\langle \phi_q^p|\phi_m^n\rangle&=&\int\,d^2R\,[\phi^{(p)}_q(\vett{r},t)]^*\phi^{(n)}_m(\vett{r},t)\nonumber\\
&=&2\pi\,\delta_{m,q}\frac{(p+n-2)!}{(\alpha+\beta-i\zeta)^{p+n}},
\earr
where we have omitted the integration along the propagation direction, to avoid the appearance of a $\delta(0)$-term, typical of scalar products of infinite-energy-carrying waves. This problem can be easily solved by using generalised, rather than fundamental, X-waves. In that case, in fact, the presence of a Laguerre polynomial in the spectrum makes the spectral index $n$ a \emph{proper} index, as it can be associated to the orthogonality relation of Laguerre polynomials \citep{book_Localized_Waves, marco_ultrashort_pulse}.

For the case of Bessel-X waves, instead, we have the following result
\barr
&&\langle\phi_n(b)|\phi_m(a)\rangle = \int\,d^2R\,[\phi_n(\vett{r},t;b)]^*\phi_m(\vett{r},t;a)\nonumber\\
&=&\frac{2\pi\delta_{n,m}}{\sin\vartheta_0}\int_0^{\infty}\,dk\,\frac{e^{-2\alpha k}}{k}\text{J}_{2m}(2a\sqrt{k})\text{J}_{2m}(2b\sqrt{k}),
\earr
whose explicit expression can be calculated in terms of Hypergeometric functions  \citep{table_of_integrals_book}. 

Despite this, the radial structure of X-waves can be controlled independently from its other spatial degrees of freedom, since it only requires the manipulation of its spectrum. Moreover, the ``non-orthogonality" of the spectral index does not affect the possibility of using this degree of freedom to encode information in both classes of X-waves. One should only be careful in designing a suitable transmission and detection protocol, able to distinguish between the different values of the spectral indices. For the case of fundamental X-waves, for example, information can be encoded by suitably shaping their spectrum with existing spectral tailoring techniques \citep{book_Ultrafast_Optics} (or simply by using generalised X-waves instead), and information can be decoded in detection by using spectral filters centered at the carrier frequency of the various transmitted X-waves, since, as it shown in Sect. III.B, the central frequency in the exponentially decaying spectrum of fundamental X-waves grows linearly with the spectral index $n$. A similar scheme could also be implemented for Bessel-X waves.
\section{Linear and Angular Momentum}
To conclude our analysis on the effect of spectral indices on ultrashort localised pulses, in this section we consider their impact on linear and angular momenta of both fundamental X-waves and X-waves with Bessel spectrum. According to standard definitions \cite{angular_momentum_of_light} their expressions are given by
\bseq\label{pj1}
\begin{align}
\vett{P} &=\frac{\varepsilon_0}{2}\int\,d^3r\, \mathcal{P}(\vett{r}),\\
\vett{J} &=\frac{\varepsilon_0}{2}\int\,d^3r\,\mathcal{J}(\vett{r}),
\end{align}
\eseq
where $\mathcal{P}$ and $\mathcal{J}$ are the linear and angular momentum densities, respectively, whose explicit expression is given by
\begin{subequations}\label{pj2}
\begin{align}
\mathcal{P}(\textbf{r}) &=\operatorname{Re}\left[\mathbf{E} (\textbf{r},t) \times \mathbf{B}^* (\textbf{r},t)\right],\\
\nonumber\\
\mathcal{J}(\textbf{r}) &=\operatorname{Re}\left[\mathbf{r} \times \mathcal{P}(\textbf{r})\right].
\end{align}
\end{subequations}
\begin{figure*}[t!]
\centering
\includegraphics[width=1\linewidth]{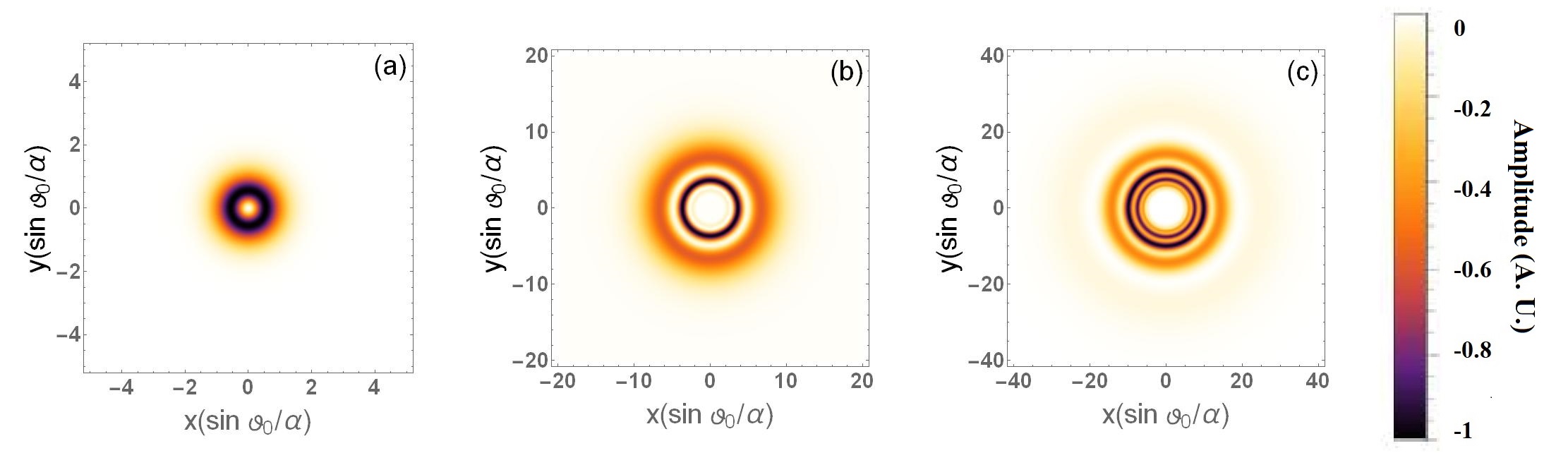}
\caption{Plot of the longitudinal component $\mathcal{L}_z(\vett{r})$ of the orbital angular momentum of X-waves with Bessel spectrum in the plane  $ \zeta=0.$ for (a) $a=1$, (b) $a=5$, and (c) $a=10$. All other parameters remain same as Fig. \ref{ER_field_bessel(m=2,a=1,5,10)}.  }\label{orbital_angular_momenta_bessel(m=2,a=1,5,10)}
\end{figure*}

Notice, that the above expressions are not only valid for monochromatic beams, but also for pulses, provided that the electric and magnetic fields are represented as analytic signals. Moreover, since we are considering paraxial pulses, we decompose the angular momentum in its orbital (OAM) and spin (SAM) components, i.e., $\mathcal{J}(\vett{r})=\mathcal{L}(\vett{r})+\mathcal{S}(\vett{r})$, where
\bseq
\begin{align}
\mathcal{L}(\vett{r}) &= \sum_{d=1}^{3}\operatorname{Re}\left[E_d(\vett{r},t)\left(\vett{r}\times\nabla\right)A_d^*(\vett{r},t)\right],\\
\mathcal{S}(\vett{r}) &= \operatorname{Re}\left[\vett{E}(\vett{r},t)\times\vett{E}^*(\vett{r},t)\right],
\end{align}
\eseq
where $d=\{1,2,3\}=\{R,\theta,z\}$, and $\vett{A}(\vett{r},t)$ is the (TM) magnetic vector potential, defined in terms of Hertz potentials as follows
\barr
\vett{A}(\vett{r},t) &=& \int\,dk\,g(k)\Bigg\{\frac{1}{c^2}\frac{\partial\vett{P}(\vett{r},t)}{\partial t}\nonumber\\
&-&\frac{i}{c\,k}\nabla\left[\nabla\cdot\vett{P}(\vett{r},t)\right]\Bigg\}.
\earr
Notice, that with the above definition, $\nabla\cdot\vett{A}=0$ is guaranteed.

For localized waves, however, the quantities in Eqs. \eqref{pj1} are, in general, divergent, due to the fact that both fundamental X-waves, and X-waves with Bessel spectrum carry infinite energy. For this reason, in this section we only consider their densities, i.e., Eqs. \eqref{pj2}, which are well-defined. Moreover, since we are considering paraxial pulses, we focus our attention on the longitudinal component (i.e., the component along the propagation direction) of both linear and angular momenta solely.

\subsection{Linear Momentum}
The longitudinal component of the linear momentum for fundamental X-waves and X-waves with Bessel spectrum for different values of the correspondent spectral indices is shown in Figs. \ref{linear_momenta_expo(m=2,n=0,5,30)} and \ref{bessel_fn_P(m=2,a=1,5,10)}, respectively. The correspondent expressions of $\mathcal{P}_z(\vett{r})$ for both cases are given as 
\barr
\mathcal{P}_z(\vett{r}) &=& E_{r}(\vett{r})B_{\theta}^*(\vett{r})-E_{\theta}(\vett{r})B_r^*(\vett{r})\nonumber\\
&=&-\frac{1}{4\cos\vartheta_0}\Bigg\{\Bigg[\text{J}_{m-1}(k\rho)-\text{J}_{m+1}(k\rho)\Bigg]^2\nonumber\\
&+&\frac{4\,m^2}{k^2\rho^2}\text{J}_m(k\rho)^2\Bigg\},
\earr
for the case of fundamental X-waves, and
\barr\label{PzBessel}
\mathcal{P}_z(\vett{r}) &=& E_{r}(\vett{r})B_{\theta}^*(\vett{r})-E_{\theta}(\vett{r})B_r^*(\vett{r})\nonumber\\
&=& -\frac{\sin^2(2\vartheta_0)}{4\cos\vartheta_0}\Bigg[\Bigg(\frac{\partial^2\phi}{\partial\xi\partial\rho}\Bigg)^2\nonumber\\
&+&\frac{m^2}{\rho^2}\Bigg(\frac{\partial\phi}{\partial\xi}\Bigg)^2\Bigg],
\earr
for X-waves with Bessel spectrum. As it can be seen, unsurprisingly, in both cases the longitudinal component of the linear momentum follows a similar structure, that
the one of the correspondent electric (magnetic) field. In this case as well, in fact, a richer spatial structure appears, in the form of extra rings. This feature could be useful, for example, to engineer novel schemes for particle control and manipulation, since the particle-field momentum exchange is sensible to the local, rather than global, momentum density. Unlike the case of fundamental X-waves, for the case of X-waves with Bessel spectrum, the transverse size of $\mathcal{P}_z(\vett{r})$ grows, as the spectral index $a$ increases. This is another manifestation of the spatio-temporal correlation. According to Fig. \ref{k_vs_gk_bessel}, in fact, as $a$ increases, the spectrum becomes more localized around smaller frequencies. In turn, therefore, the spatial distribution of the field increases in width, thus influencing the longitudinal component of the Poynting vector as well.

For the case of X-waves with Bessel spectrum depicted in Fig.  \ref{bessel_fn_P(m=2,a=1,5,10)}, we observe, from Eq. \eqref{PzBessel}, that $\mathcal{P}_z(\vett{r})\simeq\sin^2\vartheta_0$. Since we are in the paraxial regime, i.e., we assumed $\vartheta_0\ll 1$, one could be tempted to neglect this contribution and say, that the linear momentum for such pulses has negligible component along the propagation direction. However, a closer analysis reveal, that $\mathcal{P}_r(\vett{r})=0$ because of symmetry, and $\mathcal{P}_{\theta}(\vett{r})\simeq\sin^3\vartheta_0$. Therefore, $\mathcal{P}_z(\vett{r})$ represents the leading order in $\theta_0$. Notice, moreover, that from Fig. \ref{bessel_fn_P(m=2,a=1,5,10)} it is clear, that as the number of rings grows (i.e., as $a$ grows), the local size of $\mathcal{P}_z(\vett{r})$ also grows, due to the existing correlation between spectrum and space discussed above. In addition to that, the rings start assuming a pattern, that makes them narrower towards the center of the pulse, and wider towards its edges. 
\subsection{Orbital Angular Momentum}
For the OAM as well, we concentrate only on its longitudinal component, as, because of symmetry, $\mathcal{L}_r(\vett{r})=0$, and, because of paraxiality, $\mathcal{L}_{\theta}(\vett{r})\ll\mathcal{L}_z(\vett{r})$. Moreover, we only report the OAM density for X-waves with Bessel spectrum, since the one for fundamental X-waves does not contain any new information, with respect to standard results  \cite{angular_momentum_of_light}. One interesting feature that emerges from Fig. \ref{orbital_angular_momenta_bessel(m=2,a=1,5,10)} is that, as the number of rings increases, the OAM density possessed by each rings varies accordingly. 

This suggests the possibility to use X-waves with Bessel spectrum as a way to realise multi-rotational traps for dielectric particles. Particles finding themselves in different regions of the intensity distribution of the pulse, in fact, will experience a different local OAM, and, ultimately, a different scattering force, which will trigger their rotational motion (as in the case of LG-beams, for example). By suitably engineering the spectrum of X-waves with Bessel spectrum, therefore, the number of rings and their size can be controlled, thus allowing a control of the scattering force acting on each of those particle individually.

A detailed analysis of the effect of OAM on the spatial structure of both fundamental X-waves and X-waves with Bessel spectrum is reported in Appendix B.
\begin{figure*}[t!]
\centering
\includegraphics[width=1\linewidth]{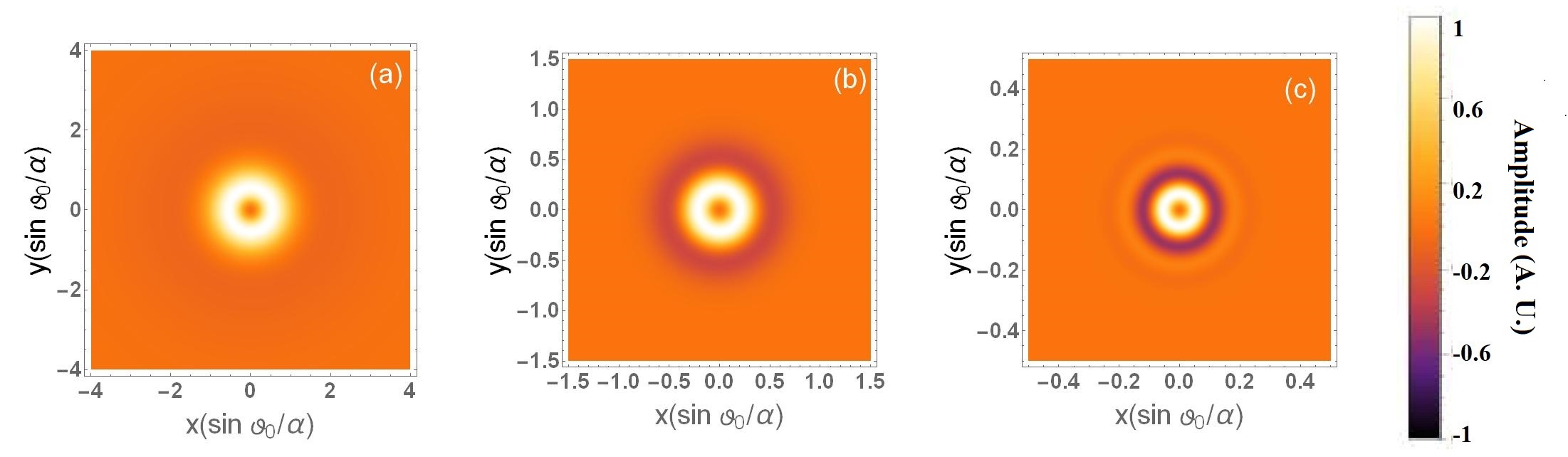}
\caption{Plot of ithe longitudinal component $\mathcal{S}_z(\vett{r})$ of the spin angular momenta of fundamental X-waves in the plane  $ \zeta=0.$ for  (a) $n=0$, (b) $n=5$, and (c) $n=30$. All other parameters remain same as Fig. \ref{EB_field_expo(m=2,n=0,5,30)}.}\label{spin_angular_momenta_expo(m=2,n=0,5,30)}
\end{figure*}
\subsection{Spin Angular Momentum}
The longitudinal component of the SAM density $\mathcal{S}_z(\vett{r})$ for fundamental X-waves and X-waves with Bessel spectrum is shown in Figs. \ref{spin_angular_momenta_expo(m=2,n=0,5,30)} and \ref{spin_angular_momenta_bessel(m=2,a=1,5,10)}, respectively. As it can be seen, in both cases, regions with positive and negative values of the SAM are appearing, revealing the presence of regions where the helicity of these pulses has different signs, i.e., a local inversion of the helicity axis occurs. Globally, however, the value of the SAM for both fundamental X-waves and X-waves with Bessel spectrum is well-determined. To understand this, notice, that the regions of negative SAM density correspond to regions of low intensity for the pulse [see, for example, Figs. \ref{EB_field_expo(m=2,n=0,5,30)} and \ref{ER_field_bessel(m=2,a=1,5,10)}].
\section{Conclusions}
In this work, we have studied the effect of the spectral index on the transverse structure of two different families of localized waves, namely fundamental X-waves and X-waves with Bessel spectrum. We have shown, that due to their innate spatio-temporal correlation, structuring the spectrum of such pulses results in the appearance of several rings in their intensity distribution, with a structure similar to those of LG-beams, for the case of fundamental X-waves.  Furthermore, we have shown how the number of rings increases in suitable steps of the spectral index, as shown Eqs. \eqref{pIndex} and \eqref{pIndexBessel} for the case of fundamental X-waves and Bessel-X waves, respectively. Within this interval, tuning of the spectral index allows a tuning of the relative intensity between these peaks. 

Our results on fundamental X-waves corroborate and generalise those given in Refs. \cite{extra1,extra2}. In these works, in fact, the ring structure of so-called Bessel-X pulses (i.e., X-waves with Gaussian spectrum) is discussed in detail, and understood as a direct consequence of the fact that the spectral bandwidth of Bessel-X pulses is narrow compared to their carrier frequency. It is well-known, that Bessel-X pulses have very similar properties to fundamental X-waves (as, in many ways, they constitute their experimentally realisable counterpart). The results obtained in our work on the role of the spectral index on the radial structure of X-waves, therefore, corroborates this statement, and finds, at the same time, confirmation in the predictions of Refs. \cite{extra1,extra2}. However, those works differ significantly from our in two important aspects: first, no contribution of OAM was taken into account (as only Bessel-X pulses of order $m=0$ were considered). Second, the analysis presented in both Ref. \cite{extra1} and \cite{extra2} only deal with scalar, rather than vector, waves. By including the influence of the OAM on the ring structure of fundamental X-waves (see Appendix B), and investigating in detail vector, rather than scalar, fields, our work extends and generalises the one reported in Refs. \cite{extra1,extra2}.

The connection of this effective radial index with the actual radial index of LG-beams has also been discussed. Despite the fact, that the newly introduced spectral index does not give rise to an orthogonality relation, we believe that it can be actively used as a novel degree of freedom, where to encode classical, and possibly quantum information. Both families of localized waves, in fact, exhibit substantial potential to be considered for higher dimensional quantum communication and quantum key distribution to encode more information which could be a potential alternative of LG beam \citep{robert_HG_LG_beam}. To this aim, an agile combination of them or combination with LG modes can raise up the performance as well \citep{angular_momentum_of_light}. The compatibility of these newly introduced spectral indices with the protocols of classical and quantum information encoding is, however, left to a future work.
\begin{figure*}[t!]
\centering
\includegraphics[width=1\linewidth]{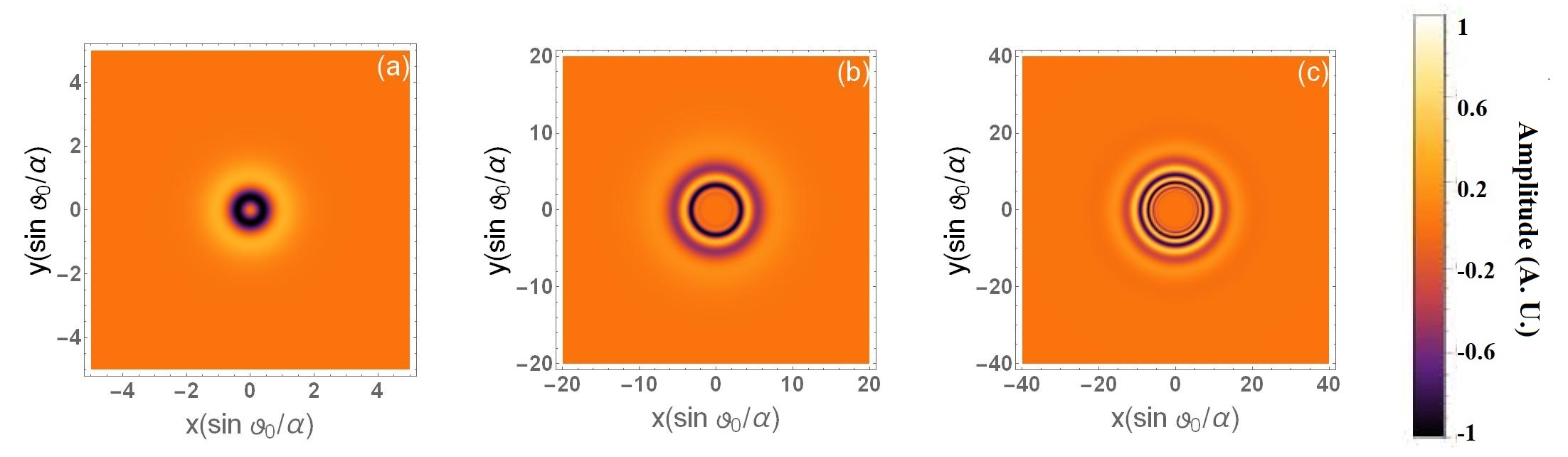}
\caption{Plot of the longitudinal component $\mathcal{S}_z(\vett{r}$ of the spin angular momenta of X-waves with Bessel spectrum in the plane $\zeta=0.$ for  (a) $a=1$, (b) $a=5$, and (c) $a=10$.  All other parameters remain same as Fig. \ref{ER_field_bessel(m=2,a=1,5,10)}. }\label{spin_angular_momenta_bessel(m=2,a=1,5,10)}
\end{figure*}
\section*{Acknowledgements}
The work is part of the Academy of Finland Flagship Programme, Photonics Research and Innovation (PREIN), decision 320165.

\section*{Appendix A: Monochromatic Vector Bessel Fields} \label{theoretical_appendix}
Using  Eqs. \eqref{TE vector eq}, we can determine the monochromatic electric fields generated by the Hertz potential given in Eq. \eqref{hertz potential monochromatic} as follows
\begin{subequations}\label{E and B field}
\begin{align}
 \mathcal{E}_R(\textbf{r},t;k) &= \frac{i}{2}\left[\text{J}_{m-1}(k\,\rho)-\text{J}_{m+1}(k\,\rho)\right]e^{i(m\theta +k\zeta)},\\
 \mathcal{E}_\theta(\textbf{r},t;k) &=-\frac{m}{k\,\rho}\text{J}_m(k\,\rho)\,e^{i(m\theta +k\zeta)},\\
 \mathcal{E}_z(\textbf{r},t;k) &=\tan\vartheta_0\,\text{J}_m(k\,\rho)\,e^{i(m\theta +k\zeta)},
\end{align}
\end{subequations}
for the (TM) electric field, and 
\bseq
\begin{align}
\mathcal{B}_R(\vett{r},t;k) &= \frac{1}{\cos\vartheta_0}E_{\theta}(\vett{r},t;k),\\
\mathcal{B}_{\theta}(\vett{r},t;k) &=-\frac{1}{\cos\vartheta_0}E_R(\vett{r},t;k),\\
\mathcal{B}_z(\vett{r},t;k) &=0,
\end{align}
\eseq
for the (TM) magnetic field. The TE electric and magnetic fields can be simply derived from the equations above by letting $\vett{E}_{TE}\rightarrow -\vett{B}_{TM}$ and $\vett{B}_{TE}\rightarrow\vett{E}_{TM}$.

\section*{Appendix B: Effect of OAM on the Transverse Intensity Distribution of Localized Pulses}
In previous works, the impact of OAM on the temporal structure of ultrashort localized pulses has been studied in detail \cite{marco_ultrashort_pulse}. There it has been shown, how an OAM-carrying ultrashort pulse self-adapts to the amount of OAM its carrying, by reducing its temporal width as the amount of OAM carried by the pulse increases. Here, we complete this picture by looking, for both fundamental X-waves and X-waves with Bessel spectrum, at the effect of OAM on the spatial structure of the pulse. The dependence on the OAM index of fundamental X-waves and X-waves with Bessel spectrum is reported in Fig. \ref{Er_field_both(m=1,5,10,n,a=5)}. As it can be seen, the behaviour is quite different for fundamental X-waves [upper panel in Fig. \ref{Er_field_both(m=1,5,10,n,a=5)}] an Bessel-X waves [lower panel in Fig. \ref{Er_field_both(m=1,5,10,n,a=5)}]. While for the former, as the amount of OAM carried by the pulse increases we observe an overall increase of the size of the intensity distribution of the electric field (i.e., its radial component), for the latter, the opposite trend is observed, namely the size of the intensity distribution shrinks as the amount of OAM carried by X-waves with Bessel spectrum is increased. This, ultimately, is another manifestation of the spatio-temporal correlation of localized waves.

\nocite{*} 
\bibliographystyle{aipnum-cp}%
%\bibliography{bibliography}%
%
\begin{figure*}[t!]
\centering
\includegraphics[width=1\linewidth]{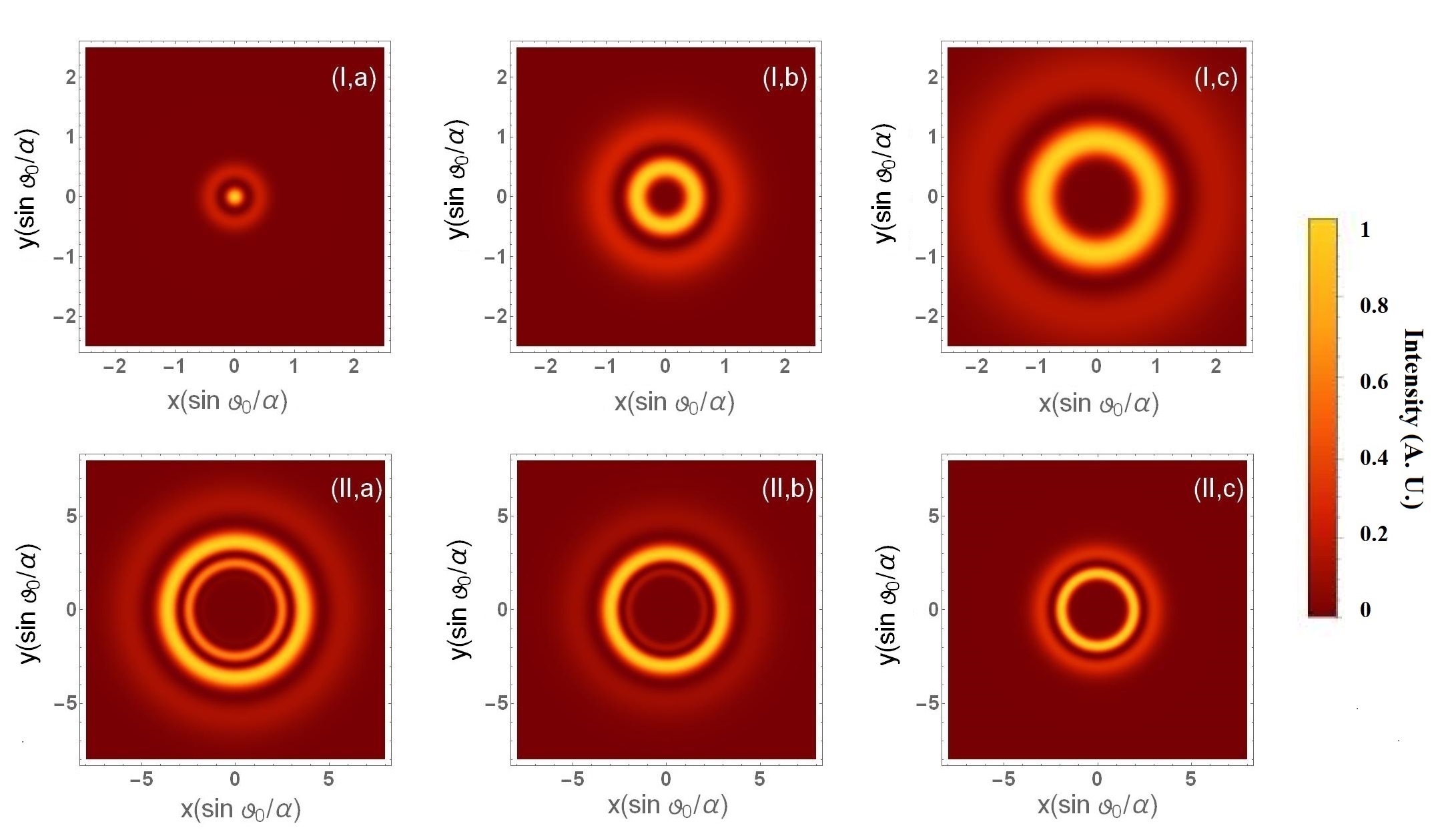}
\caption{ Plot of intensity distribution of the radial component of the electric field $\mathbf{E}_r(\vett{r,}t)$ for (I) fundamental X waves with $n=5$, and (II) X-waves with Bessel spectrum with $a=5$ in the plane $\zeta = 0$, for different OAM values, i.e., (a) $m=1$, (b) $m=5$, and (c) $m=10$. Notice, that while for fundamental X-waves (upper panel) the size of $E_r(\vett{r},t)$ increases with increasing OAM , for Bessel-X waves (lower panel) the trend is opposite, as the size of $E_r(\vett{r},t)$ decreases as the amount of OAM carried by the pulse increases. To realise these plots we have assumed $\vartheta_0=0.01$ and $\alpha=1$.} \label{Er_field_both(m=1,5,10,n,a=5)}
\end{figure*}

\bibliography{Radial_Structure}%
\end{document}